# A New Scheme for Image Compression and Encryption Using ECIES, Henon Map, and AEGAN

Mahdi Shariatzadeh, Mahdi Eftekhari, Mohammad Javad Rostami

Shahid Bahonar University of Kerman, Kerman, Iran

*Abstract*- **Providing security in the transmission of images and other multimedia data has become one of the most important scientific and practical issues. In this paper, a method for compressing and encryption images is proposed, which can safely transmit images in low-bandwidth data transmission channels. At first, using the autoencoding generative adversarial network (AEGAN) model, the images are mapped to a vector in the latent space with low dimensions. In the next step, the obtained vector is encrypted using public key encryption methods. In the proposed method, Henon chaotic map is used for permutation, which makes information transfer more secure. To evaluate the results of the proposed scheme, three criteria SSIM, PSNR, and execution time have been used.**

## I. Introduction

With the rapid expansion of communication platforms, the transmission of multimedia information has increased. Meanwhile, images and their security are very important. Image encryption and the proposal of new image encryption techniques have a special value. Artificial intelligence has also found diverse and extensive applications to solve many problems, including creating secure systems [1].

In general, cryptography is done in two ways: symmetric cryptography and public key cryptography. In symmetric cryptography, both parties - sender and receiver - use the same key for cryptography. The processes of encryption and decryption are the reverse of each other [2]. But in public key cryptography, the sender encrypts the information with the receiver's public key, and the receiver decrypts the encrypted information using their private keys [3].

In this paper, related works will be reviewed in part II. In part III, the preliminaries for the proposed method will be explained. In part IV, the proposed method will be explained. In part V, the results of the implementation of the proposed method will be presented on different images, and in part VI, the conclusion will be made.

## II. Related Works

Suhail and Sankar [4] proposed a system for image compression and encryption to securely transmit image data with minimum bandwidth. Their proposed system used autoencoder for compression and logistic chaos mapping for encryption. The main drawback of the system proposed by them is not being resistant to differential attacks. The results obtained from the implementation of their proposed method are different from the results reported by them.

Duan et al. [5] proposed an image encryption method using variational autoencoder. In their proposed method, chaos maps were not used and compression and encryption operations are based on model training. They used MSE and PSNR to evaluate their method.

Lu et al. [6] proposed a content-adaptive image compression and encryption method based on Compressed Sensing (CS). Their method worked in two steps. In the first step, they used a 4D laser chaos system to construct a measurement matrix for image compression. In the second step, they used an Improved Fractal Sorting Matrix (IFSM).

Huang and Cheng [7] proposed a method that utilized a hyperchaotic system (HCS) and discrete cosine transform (DCT). Their scheme had two stages. At first, they transformed plain images by cosine transformation into blocks. Then they permutated each block by Arnold transformation to gain a local scrambling effect.

Another algorithm was proposed by Nan et al. [8]. Their algorithm compressed and encrypted images based on block-compressed sensing, multiple S-Boxes, and a novel hyper chaos system. They also proposed a hyperchaotic system involving a 2D Logistic coupling Cubic map which is coupled with a Cubic map and Logistic map.

Gupta and Vijay [9] proposed a scheme that employed a stacked auto-encoder and logistic map for image compression and encryption. They used a seven-layer network and the backpropagation algorithm for compressing the images. Their algorithm's key was managed by the logistic chaotic map.

Ahmad and Shin [10] proposed a hybrid image compression-encryption scheme that compressed images in the encryption domain. Their encryption phase employed Chaos theory. They performed compression on the shuffled image. Then the substitution phase had some 8-bit outputs. Their lossless nature scheme was suitable for medical image compression and encryption applications.

Liu et al. [11] proposed an image compression–encryption scheme that utilized the hyperchaotic system and 2D compressive sensing. Their scheme had two steps. First, they constructed a hyperchaotic system for the measurement matrix. Then, they used 2D compressive sensing for compressing the image.

Sneha et al. [12] proposed an algorithm for image encryption using Walsh–Hadamard transform. In their algorithm, the images are processed channel-wise. for increasing the confusion, they used Arnold and Tent maps. They have claimed that the security and the key space of their image encryption system were increased by complex behaviors and random chaotic ranges of chaotic maps.

## III. PRELIMINARIES

### A. Henon chaotic map

Henon map is considered a chaotic map in the study of dynamic systems [13]. This map makes point $(x_n, y_n)$ correspond to point $(x_{n+1}, y_{n+1})$ in the coordinate plane and uses the following relationship for this purpose:

$$\begin{cases} x_{n+1} = 1 - ax_n^2 + y_n \\ y_{n+1} = bx_n \end{cases} \quad (1)$$

As can be seen in equation (1), the Henon map depends on two parameters $a$ and $b$, which for the classical Henon map have values of $a = 1.4$ and $b = 0.3$. In the classic case, this map is considered a chaotic map. For other values of $a$ and $b$, this map may exhibit other behaviors. These behaviors are chaotic, periodic, or converge to a value. Figure 1 shows the path of the Henon chaotic map.

### B. ECIES

Elliptic curve cryptography (ECC) is a public key cryptography technique based on the algebraic structure of elliptic curves over finite fields that can be used to create faster, smaller, and more efficient cryptographic keys [14]. ECC is an alternative to the RSA cryptographic algorithm and is often used for digital signatures in cryptocurrencies such as Bitcoin and Ethereum, as well as one-way encryption of emails, data, and software [15]. Other uses of elliptic curves include their indirect use in symmetric cryptography schemes as well as in several integer factorization algorithms based on elliptic curves that have applications in cryptography [16].

Elliptic curve integrated encryption scheme (ECIES) is a hybrid encryption system proposed by Victor Shoup in 2001 [17]. ECIES is standardized in ANSI X9.63, IEEE 1363a, ISO/IEC 18033-2 and SECG SEC-1 [18]. ECIES combines a key encapsulation mechanism (KEM) with a data encapsulation mechanism (DEM) [19]. The system independently derives a bulk encryption key and another key from the shared secret [20]. The output of the encryption function is the tuple $\{K, C, T\}$, where $K$ is the encrypted common secret, $C$ is the ciphertext, and $T$ is the authentication tag [21].

### C. AEGAN

Autoencoders are a class of self-supervised neural networks that learn a domain-specific encoding from a sample space. Autoencoders consist of two networks: an encoder ($E$) that learns $E: X \to Z$ ($Z$ is the value of the latent space with lower dimensions); and a decoder ($G$) that learns $G: Z \to X$. The autoencoder is tasked with encoding and then reconstructing each instance of the dataset so that $G(E(x)) = \tilde{x} \approx x$, which must be reconstructed by a loss function. This loss function for reconstruction is usually a function that calculates the pixel difference between input $x$ and output $\tilde{x}$. After training the autoencoder, it can be used for other purposes such as feature extraction [22].

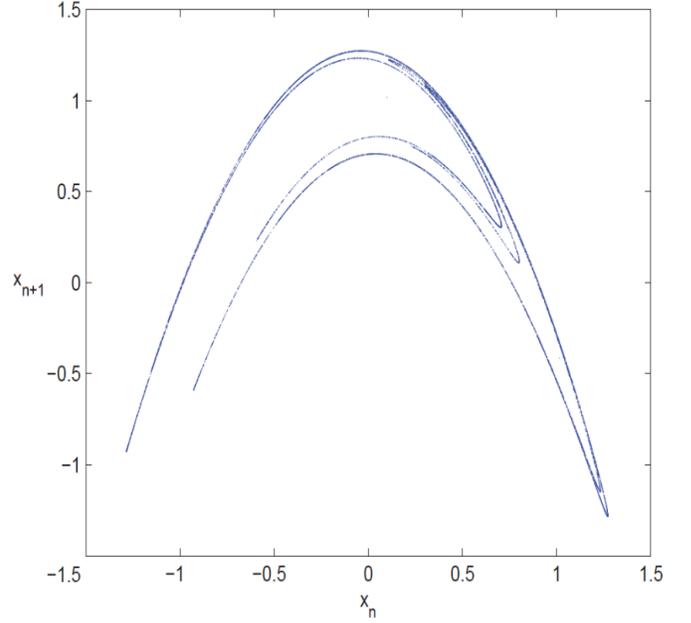

*Figure 1 Path of the Henon chaotic map.*

Generative adversarial networks (GANs) are a class of generative and unsupervised neural networks that can approximate the distribution of data to generate a given dataset. These networks consist of two networks, which are a generator ($G$) and a discriminator ($D_x$). The generator learns the mapping $G: Z \to X$. The discriminator is also tasked with distinguishing which samples are from the true distribution (i.e., the dataset) and which samples are generated by $G$. The generator and the discriminator are updated in turn in a minimax game, such that the generator tries to minimize equation (2), while the discriminator tries to maximize it [22].

$$\min_G \max_D \left[ \mathbb{E}_{x \sim p_{\text{data}}} [\log D_x(x)] + \mathbb{E}_{z \sim p(z)} [\log(1 - D_x(G(z)))] \right]. \quad (2)$$

AEGAN is a technique for learning two-way mapping between some sample space $X$ and latent space $Z$. An autoencoder network $E$ is trained to learn the function $E: X \to Z$ (mapping each real instance to a point in the latent space). The generator network $G$ is trained to learn the function $G: Z \to X$ (mapping each point in the latent space to an instance in the sample space). These networks are successively trained with two discriminator networks $D_x$ and $D_z$. $D_x$ is used to distinguish between real and generated samples. $D_z$ is used to distinguish between real and generated latent spaces [22].

Figure 2 shows the architecture of an AEGAN network. In this figure, four networks are shown as squares, data are shown as circles, and loss functions are shown as rhombuses. The colors represent the constituent models: red represents the generator for images, blue represents the autoencoder for the images, yellow represents the generator for latent spaces, and green represents the autoencoder for the latent spaces [22].

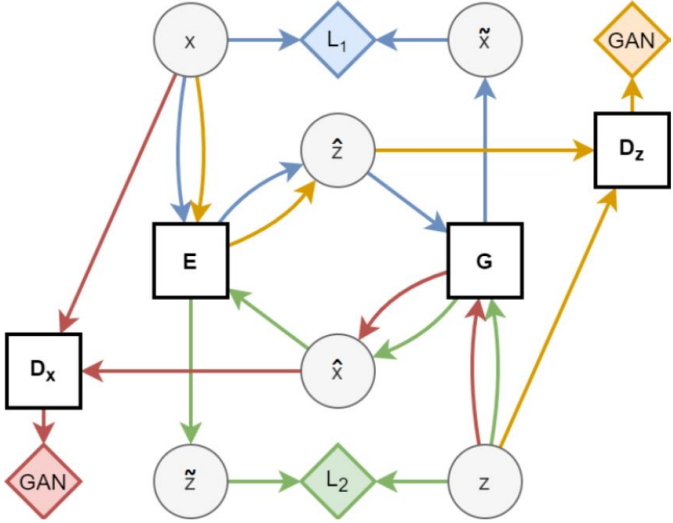

*Figure 2 The architecture of an AEGAN network [22].*

## IV. PROPOSED METHOD

The proposed method tries to compress the images by using the autoencoder part of an AEGAN network. Image compression is such that an image with dimensions $n \times n$ is placed in a one-dimensional vector of size $m$; so that m is much smaller than $n \times n$ ($m \ll n \times n$). The compression operation is done so that a large image can be transmitted in a low bandwidth. After compressing the image, using Henon chaotic map, the obtained vector is scrambled in order to increase the confusion and diffusion in the encryption system. In the last encryption step, the scrambled vector is encrypted using ECIES. The output of the encryption part is a vector of size $m$. Encryption is done using a public key scheme. Therefore, in the proposed method, there are three keys to perform the encryption and decryption process. Two keys for ECIES which include public key (PUB_KEY) and private key (PRIV_KEY). The third key is used to specify an initial value for Henon chaotic map. The third key is a symmetric key (SYM_KEY).

In the decryption process, an attempt is made to reconstruct the desired image using the encrypted vector. First, ECIES decrypts the encrypted vector using the PRIV_KEY. The result of this step is the vector that is scrambled in the encryption step. The scrambled vector is unscrambled using SYM_KEY and the initial value obtained for the Henon chaotic map. The result of this step is the vector created in the encryption step by the autoencoder from the AEGAN network. This vector is the compressed image. In order to reconstruct the image in the final stage of decryption, it is necessary to send the compressed image to the generator part of the AEGAN network.

The mentioned generator can reconstruct the desired image by using the received vector. In our experiments, the AEGAN network is trained using the DIV2K dataset on 800 images of $256 \times 256$ dimensions [23]. The size of the latent space for the AEGAN network is considered to be 100 ($m = 100$). The two trained models - meaning the GAN and the autoencoder - are used for compression in encryption and for image reconstruction in decryption.

### A. Compression and encryption operations

In the proposed method, the compression and encryption operations are performed in three steps. The input of this operation will be an image and its output will be an encrypted feature vector. The ratio of input and output dimensions of this method is different from each other.

In the first step, the input image will be converted into a feature vector using the autoencoder part of an AEGAN network. For this method, the dimension of the latent space is 100. Therefore, in the first step, an image with dimensions of $256 \times 256$ is mapped to a vector of 100. It should be noted that the AEGAN network used in this method is a network trained using the DIV2K dataset and on 800 images with dimensions of $256 \times 256$. The size of the latent space for the mentioned network is a vector of 100.

In the second step, the feature vector obtained from the first step is shuffled. At this stage, a pseudo-random sequence is created using the Henon chaotic map. As mentioned before, in this method, three keys are used to encrypt images. The symmetric cryptographic key (SYM_KEY) is the coordinates of a point $(x_0, y_0)$ on the plane, which is generated using equation (1) of the coordinates of 100 other points. By considering the first element of the generated points, a pseudo-random sequence is obtained, and by sorting the sequence and considering the index of the elements, the feature vector can be scrambled up.

In the third step, using the elliptic curve integrated encryptor and using the public key (PUB_KEY), the scrambled feature vector is encrypted. The encrypted output will be a vector with the dimensions of the feature vector ($m = 100$). The pseudocode of the encryption process of the proposed method is shown in Algorithm 1.

Algorithm 1. Compression and encryption process of the proposed method.

| |
|---|
| Input: InputImage, $SYM\_KEY, PUB\_KEY$ |
| Output: EncryptedVector |
| 1: State ← InputImage |
| 2: vector ← AutoEncoder ( State ) |
| 3: ChaoticSequence ← HenonMap ($SYM\_KEY$) |
| 4: vector ← Shuffler (vector, ChaoticSequence ) |
| 5: vector ← ECIES_ Encrypt ( Vector, PUB_KEY) |
| 6: EncryptedVector ← vector |
| 7: return EncryptedVector |

### B. Extraction and decryption operations

The decryption process is performed in three steps just like the encryption operation. The input of this operation will be an encrypted vector and the output will be a reconstructed image. In the first step, the encrypted vector is decrypted using an integrated elliptic curve decryptor and with the help of a private key (PRIV-KEY). The decrypted output is a shuffled vector of

100 which is required to be de-shuffled in the second step. In the second stage of decryption, the reverse of the corresponding operations in encryption is performed.

At this stage, a pseudo-random sequence is created using SYM_KEY -which is the coordinates of a point on the plane- and with the help of Henon chaotic map. By sorting the generated sequence and considering the index of the elements, the vector of the previous step can be de-shuffled.

In the third step, using the generator part of AEGAN network and with the help of the vector obtained from the previous step, the desired image can be reconstructed. The results of the decryption process are presented in Section V. Also, the pseudocode of the extraction and decryption process of the proposed method is shown in Algorithm 2.

Algorithm 2. Extraction and decryption process of the proposed method.

Input: EncryptedVector, SYM_KEY, PRIV_KEY
Output: ReconstructedImage
1: Vector ← EncryptedVector
2: Vector ← ECIES_Decrypt(Vector, PRIV_KEY)
3: ChaoticSequence ← HenonMap (SYM_KEY)
4: Vector ← Deshuffler (Vector, ChaoticSequence )
5: State ← GAN(Vector)
6: ReconstructedImage ← State
7: return ReconstructedImage

## V. Experimental Results

In the proposed method, the size of the encrypted vector is 100. This means that the proposed method compresses the size of the input images from 65536 pixels to 100 pixels (assuming that each element of the output vector is one pixel). Since the number of pixels is very small and also less than 255, obtaining the values of some evaluation criteria such as entropy, correlation coefficient, etc. does not help to evaluate this method because these values are small. For example, in the case of entropy, which is a measure of the randomness of pixels in an image, the pixel values of the encrypted image should be uniformly distributed from 0 to 255 and cover the entire image area to hide any patterns in the final output. In the proposed method, the final encrypted image with a size of 100 pixels cannot contain all values from 0 to 255, resulting in a low entropy value (less than 8). Therefore, it is not reasonable to compare the entropy value of this method with other entropy values obtained by other methods. Since the encrypted image is very small (a 100-size vector), most metrics have lower values than those obtained by other methods. Compared to other methods, at least 255 pixels are required in the final encrypted image. The proposed method focuses on transmitting encrypted images with very low bandwidth. In the following, the results of calculating SSIM, PSNR and execution time for the proposed method are presented.

### A. SSIM

The Structural Similarity Index (SSIM) is a perceptual metric that measures the degradation of image quality caused by processes such as data compression or loss in data transmission [24]. This criterion requires two images from a captured image (a reference image and a processed image). The processed image is usually compressed. SSIM is best known in the video industry but has many applications in still photography, colorization of black and white images, image steganography, and image encryption [25]. SSIM actually measures the perceptual difference between two similar images [26]. In fact, it is impossible to judge which of the images is better. This should be inferred from knowing which one is the original image and which image has undergone additional processing such as data compression. SSIM is based on visible structures in the image. Structural similarity index measurement (SSIM) is a method for predicting the perceived quality of digital television and motion images, as well as other types of digital images and videos [27]. This criterion is calculated as follows:

$$\text{SSIM}(x,y) = \frac{(2\mu_x * \mu_y + C_1)(2\sigma_{xy} + C_2)}{(\mu_x^2 + \mu_y^2 + C_1)(\sigma_x^2 + \sigma_y^2 + C_2)}. \quad (3)$$

In equation (3), $\mu_x$ and $\mu_y$ mean the mean of $x$ and $y$, respectively. $\sigma_x^2$ and $\sigma_y^2$ are the variances of $x$ and $y$, respectively, and $\sigma_{xy}$ is the covariance of $x$ with $y$. In the equation (3), the values of $C_1$ and $C_2$ are calculated as follows:

$$C_1 = (k_1 L)^2, C_2 = (k_2 L)^2 \quad (4)$$

The variable $L$ is the dynamic range of pixel values. The default values of $k_1$ and $k_2$ are as follows:

$$k_1 = 0.01, k_2 = 0.03. \quad (5)$$

As mentioned, structural similarity index is a method to measure the similarity between two images. Here, the input image and the reconstructed image at the receiver side are considered for SSIM calculation. In Table 1, SSIM results are calculated and presented based on relations (3), (4), and (5) for several images.

### B. PSNR

Peak signal-to-noise ratio (PSNR) is an engineering term for the ratio between the maximum possible power of a signal and the power of disruptive noise that affects its display accuracy [28]. Since many signals have a very wide dynamic range, PSNR is usually expressed as a logarithmic quantity using a decibel scale [29]. PSNR is commonly used to quantify the quality of reconstruction of images and videos subjected to compression. PSNR is used as a quality measure between original and decoded images. PSNR can be calculated as follows:

$$PSNR = 10 \, log_{10} \frac{(2^n - 1)^2}{MSE}. \quad (6)$$

In equation (6), $n$ means the number of bits per pixel. $MSE$ is the mean squared error between the pixels of the original image and the decrypted image, which is calculated as follows:

$$MSE = \frac{1}{MN} \sum_{x=1}^{x=M} \sum_{y=1}^{y=N} [O(x,y) - R(x,y)]^2, \quad (7)$$

In equation (7), x and y are pixel coordinates of the image with size $M \times N$. $O$ and $R$ are the original and decrypted images, respectively. The range of $MSE$ is $[0, +\infty)$. The $MSE$ value between the original and decoded image should be minimum. In Table 1, PSNR results are calculated and presented based on equation (6) for several images. Values higher than 27 dB for the PSNR measure indicate a suitable reconstruction power for the proposed method. Although these obtained values are different from the ideal value for PSNR, which is infinite, it can have applications in systems with low bandwidth.

*C. Execution Time*

Execution time refers to the time required to execute a particular image encoding technique. This time is actually the compilation time and the execution time. The execution time should be minimal to perform the image encryption process. Execution time is generally measured in seconds, milliseconds, or minutes [30].

Considering that in the proposed method, the process of compression and encryption is different from the process of extraction and decoding, in Table 1, the results of the execution time by encryption decryption are calculated and presented in seconds for several images.

## VI. CONCLUSION

In this paper, a method for compressing and encryption images was proposed. It was shown that images can be recovered by using generative models such as AEGAN. The proposed method works in such a way that low bandwidth channels can be used to transmit information safely. The combined use of the Henon chaotic map and public key encryption made the proposed method have the advantages of symmetric encryption and public key encryption. The results of reconstructing encrypted images showed that the proposed method can have practical applications.

*Table 1 The results of the proposed method on different images.*

| Input image | Restored image | SSIM | PSNR (dB) | Encryption time (sec) | Decryption time (sec) |
|---|---|---|---|---|---|
| 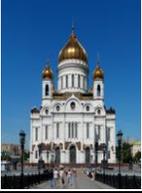 | 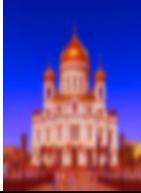 | 0.4254 | 27.736 | 0.1933 | 0.3718 |
| 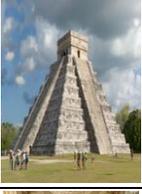 | 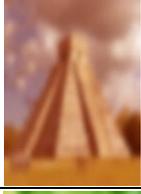 | 0.5336 | 27.524 | 0.1081 | 0.3186 |
| 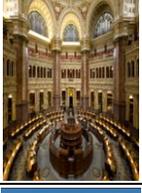 | 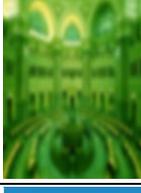 | 0.2489 | 27.849 | 0.1767 | 0.4058 |
| 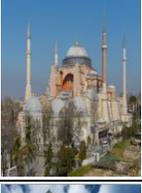 | 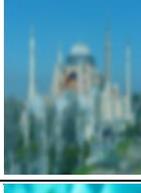 | 0.5868 | 27.749 | 0.1435 | 0.3475 |
| 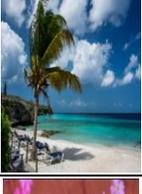 | 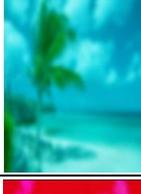 | 0.4214 | 27.825 | 0.2973 | 0.4416 |
| 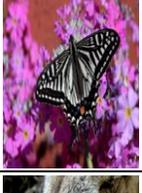 | 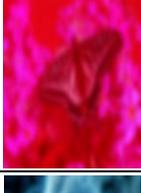 | 0.2452 | 27.902 | 0.2011 | 0.3348 |
| 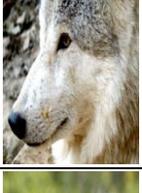 | 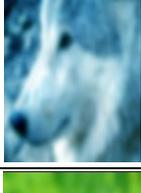 | 0.3501 | 28.062 | 0.2058 | 0.4091 |
| 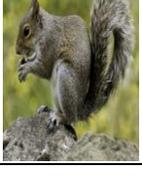 | 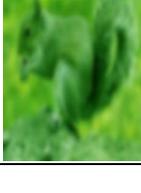 | 0.4784 | 27.461 | 0.1973 | 0.3624 |